Title: Conceivable helical form of light propagation may signify symmetry-breaking similar to Jahn-Teller effects
Author: Mladen Georgiev (Institute of Solid State Physics, Bulgarian Academy of Sciences, 1784 Sofia, Bulgaria)
Comments: 9 pages including 3 figures, all pdf format
Subj-class: physics

We stress on the similarities between vibronic symmetry-breaking, due to the mixing of electronic states by phonons and leading to the appearance of vibronic polarons along a helix, and the conceivable helical form of light propagation, possibly arising from a similar symmetry- breaking due to the mixing of fermion states by photons.

1. Introduction

This study has been incited by a paper kindly made available by its author [1]. He has raised a curious question as to whether light propagates along a straight line in a rectilinear motion or moves along a helical path combining rotation with translation. He has presented arguments to support his view, though the strongest doubts have been cast because of a numerical estimate suggesting that it will be all but impossible to check the phenomenon experimentally because of the desperately small photon mass.

Being engaged for quite a bit of time (45 years now) with research in physics, in particular symmetry lowering effects in matter due to distorting interactions, I gradually became aware that there should be an universal symmetry-breaking coupling in nature operative in both condensed and nuclear matter. Gradually, that idea crystallized to the point of becoming a likely version. Consider the Jahn-Teller or, better, vibronic coupling and generalize the latter to comprise the mixing of nearly-degenerate fermion states through coupling to an appropriate boson mode. Now, count the various possibilities of basic fermion types (electrons and holes, nucleons) and boson fields (phonon, photon, π-meson, etc.) and you will get an impressive number of symmetry-breaking options where a higher symmetry is destroyed by virtue of the coupling of fermion state pairs to bosons, taking account of a bunch of couplings known in chemistry as vibronic-, Jahn-Teller-, Renner- effects, etc.

Well, but what is the link to the propagation of light? Seems to be logically simple. Consider the following scenario: during its propagation across a material medium a beam of light couples to mix vibronically nearly-degenerate electron or hole states which destroys inversion symmetry along the way. An obvious result of the loss of parity is the appearance of the spiral whereas helical trajectory replaces the rectilinear one which conserves parity. But, what if the hypothesis is not subjet to proof due to the negligible photon mass? Well, there seems to be a way our, at least for the time being: the vibronic coupling in condensed matter gives rise to a specific type of vibronic polaron, plays a basic role in a number of phenomena in solids and very likely moves helically due to the broken inversion symmetry in much the same way light may be expected to move. If so, the phenomenon might be studied on the vibronic polaron scale before an experimental technique is devised for studying the propagation of light.

The paper is further organized as follows: After a brief concise survey of what is available on the spirals, the vibronic physics will be described in some detail, not too much because of a wealth of literature already published. Then the premises of itinerant vibronic polarons will be outlined aimed at proving the natural appearance of a helix to replace, due to the breakup of inversion symmetry, the rectilinear propagation.

## 2. Helices

### 2.1. Mathematics

The helical form of light propagation has long been the subject of debate [1] with the result that at the present time we still do not know much about its intimate mechanism. Classically speaking, the motion along a spiral may be represented as an uniform rectilinear one, say along the z-axis, complemented by two oscillators along the x- and y- axes vibrating at the same frequency. We define the following parametric equations in rectangular coordinates, where $r_\varphi$ is the planar radius $r_\varphi$:

$x = r_\varphi \cos(\omega t)$

$y = r_\varphi \sin(\omega t)$

$z = r_\varphi \cotan\theta = vt$ (1)

where $\theta$ is the helical angle. The polar radius obtains as:

$r_\varphi = \sqrt{(x^2 + y^2)}$ (2)

Using (2) the spherical radius meets:

$r_{\varphi,\theta} = \sqrt{(x^2 + y^2 + z^2)} = = \sqrt{[r_\varphi^2 + (vt)^2]}$, (3)

and we also get.

$\cotan\theta = (vt/r_\varphi)$ $(\infty \leq vt \leq r_\varphi)$

$\theta = \cotan^{-1}(vt/r_\varphi)$ $(0 \leq \theta \leq \frac{1}{4}\pi)$ (4)

The *general helix*, is expressed in rectangular coordinates as $t \rightarrow (a\cos t, a\sin t, bt)$, $t \in [0,T]$ for $a = r_\varphi$, $t = \omega t$, $bt = vt$, $b = v/\omega$. Its *length* is $T\sqrt{(a^2 + b^2)} = 2\pi\sqrt{[r_\varphi^2 + (v/\omega)^2]}$, its *curvature* is $|a|/(a^2 + b^2) = |r_\varphi|/[r_\varphi^2 + (v/\omega)^2]$, its *torsion* is $b/(a^2 + b^2) = (v/\omega)/[r_\varphi^2 + (v/\omega)^2]$ [2].

In mathematics, a helix is a curve in 3-dimensional space. The following three equations in rectangular coordinates define a simple helix form: $x = \cos t$, $y = \sin t$, $z = t$. As the parameter *t* increases, the point (*x,y,z*) traces a right-handed helix of pitch $2\pi$ about the *z*-axis, in a right-handed coordinate system.

In cylindrical coordinates ($r$, $\theta$, $h$), the same helix is described by $r = 1$, $\theta = t$, $h = t$.

Another way of mathematically constructing a helix is to plot a complex valued exponential function ($e^{ix}$) taking imaginary arguments (see Euler's formula).

## 2.2. Helical types

A helix (pl: helices), from the Greek word *ἑλικας/ἕλιξ*, is a three-dimensional, twisted shape. Common objects formed like a helix are a spring, a screw, and a spiral staircase (though the last would be more correctly called helical). Helices are important in biology, as the DNA molecule is formed as two intertwined helices, and many proteins have helical substructures, known as alpha helices.

Helices can be either right-handed or left-handed. With the line of sight being the helical axis, if clockwise movement of the helix corresponds to axial movement away from the observer, then it is a right-handed helix. If counter-clockwise movement corresponds to axial movement away from the observer, it is a left-handed helix. Handedness (or chirality) is a property of the helix, not of the perspective: a right-handed helix cannot be turned or flipped to look like a left-handed one unless it is viewed through a mirror, and vice versa.

Most hardware screws are right-handed helices. The alpha helix in biology as well as the A and B forms of DNA are also right-handed helices. The Z form of DNA is left-handed.

Except for rotations, translations, and changes of scale, all right-handed helices are equivalent to the helix defined above. The equivalent left-handed helix can be constructed in a number of ways, the simplest being to negate either the x, y or z component.

A double helix typically consists geometrically of two congruent helices with the same axis, differing by a translation along the axis, which may or may not be half-way.

A conic helix may be defined as a spiral on a conic surface, with the distance to the apex an exponential function of the angle indicating direction from the axis. An example of a helix would be the Corkscrew roller coaster at Cedar Point amusement park.

A circular helix has constant curvature and constant torsion. The pitch of a helix is the width of one complete helix turn, measured along the helix axis.

## 2.3. Quantum mechanics

The quantum-mechanical helix is reduced to two linear harmonic oscillators each along the x- or y- axes and a plane wave propagating along the z-axis. The three corresponding events being independent of each other, the complete steady-state wave function of the system is composed of a product, e.g.:

$$\Psi(\mathbf{r}) = \psi_x(\mathbf{x})\psi_y(\mathbf{y})A\exp(ikz) \tag{5}$$

with

$$E = E_x + E_y + E_z = h\nu_x(n_x + \tfrac{1}{2}) + h\nu_y(n_y + \tfrac{1}{2}) + h^2k^2/4\pi m \tag{6}$$

standing for the complete energy eigenvalue, and in vibrational state $\psi_x(\xi) = H_n(\xi)\exp(-\tfrac{1}{2}\xi^2)$ where $\xi = \alpha x$, etc. [3].

Attaching chirality $\chi$ to a helix, e.g. $\chi = 1$ (right-handedness) and $\chi = -1$ (left-handedness), the right-to-left transition changes chirality $|\Delta\chi| = 2$, the right-to-right (left-to-left) transition occurs at constant chirality $\Delta\chi = 0$. We can also extend the definition by adding up zero-chirality $\chi = 0$ which will be exhibited by the rectilinear path of light propagation ($r_\varphi = 0$). It will be of particular interest to qualify transitions from $\chi = 0$ to $\chi = 1$ or to $\chi = -1$. These may be regarded as non-conserving chirality transitions ($|\Delta\chi| = 1$). Our further goal will be tracing the origin of these transitions, whether they do not arise from the breaking of symmetry similar to the vibronic effects.

These interconversions can be expressed in terms of the absolute chirality, as follows: $\Delta|\chi| = 0$ from one handedness to another one and $\Delta|\chi| = 1$ between a handedness to a non-handedness. The former ones do conserve chirality, the latter one does not. The further question is just what causes the loss or initiation of chirality in the latter processes. It reduces to what causes the loss or initiation of curvature. This can be expressed in terms of the polar radius $r_\varphi \neq 0$ (handedness) and $r_\varphi = 0$ (non-handedness).

### 3. Hamiltonian for off-center (vibronic) polarons

The three chirality cases of $\chi = \pm 1, 0$ correspond to three potential energy curves, as shown in Figure 1. These are vibronic potential energy curves which arise on solving for the oscillatory part of the wave functions in (5) and represent the potentials of two displaced oscillators for $\chi = \pm 1$ and one non-displaced oscillator at $\chi = 0$. The displacement along the oscillator coordinate comes by virtue of a force arising from the mixing of two opposite-parity nearly-degenerate fermion states by an odd-parity boson mode. Examples of boson fields coupling to fermions are not abundant though they are signifying: e.g. phonons and photons in soft or hard condensed matter. The starting Hamiltonian reads:

$$H = \sum_{m\alpha} E_{m\alpha} a_{m\alpha}^\dagger a_{m\alpha} + \sum_{i\alpha\beta} G_{i\alpha\beta} q_{i\alpha\beta} (a_{m\alpha}^\dagger a_{m\beta} + a_{m\beta}^\dagger a_{m\alpha}) +$$

$$\tfrac{1}{2}\sum_{i\alpha\beta} K_{i\alpha\beta} q_{i\alpha\beta}^2 + \tfrac{1}{2}\sum_{i\alpha\beta} (\eta^2/m_{i\alpha\beta})(\partial^2/\partial q_{i\alpha\beta}^2) \qquad (7)$$

where $a_{m\alpha}$, etc are fermion ladder operators, $q_{i\alpha\beta}$ are the boson coordinates, $K_{i\alpha\beta}$ is the stiffness, $G_{i\alpha\beta}$ is the boson-fermion coupling constant, $\eta = h/2\pi$. At this stage $q_{i\alpha\beta}$ is a c-number but on quantizing it through introducing boson ladder operators by means of $(m_{i\alpha\beta}\omega_{i\alpha\beta}/\eta)^{1/2} q_{i\alpha\beta} = (b_{i\alpha\beta}^\dagger + b_{i\alpha\beta})$, we obtain instead:

$$H = \sum_{m\alpha} E_{m\alpha} a_{m\alpha}^\dagger a_{m\alpha} + \sum_{i\alpha\beta} G_{i\alpha\beta} (m_{i\alpha\beta}\omega_{i\alpha\beta}/\eta)^{-1/2} (b_{i\alpha\beta}^\dagger + b_{i\alpha\beta})(a_{m\alpha}^\dagger a_{m\beta} + a_{m\beta}^\dagger a_{m\alpha}) +$$

$$\sum_{i\alpha\beta} (n_{i\alpha\beta} + \tfrac{1}{2})\eta\omega_{i\alpha\beta} b_{i\alpha\beta}^\dagger b_{i\alpha\beta} \qquad (8)$$

Hamiltonian (7) generates off-center vibronic polarons by the adiabatic approximation which discards the lattice kinetic energy ($\partial^2/\partial q_{i\alpha\beta}^2 \equiv 0$) on solving for the electronic energies whereas the remaining lattice quantities are ascribed the role of parameters.. Once this has been done, the vibronic part of the problem is dealt with as one in which the electronic eigenvalues and lattice-dependent parametric terms all serve the role of vibronic potential energy, while the lattice kinetic energy term is rehabilitated. The resulting vibronic equation describes the quantum-mechanical behavior of vibronic polarons whose main feature is the breakup of spatial inversion. At this stage the solution to the (7)-based Schrödinger equation tells of the vibrational structure, binding energy and polarization of the vibronic polaron band [9].

The adiabatic procedure (two-site two-level approach) gives rise to a secular equation for $q_{i\alpha\beta}$

$$E^2 - (H_{11} + H_{22})E - H_{11}H_{22} - [\Sigma_{i\alpha\beta}G_{i\alpha\beta}q_{i\alpha\beta}(a_{m\alpha}^{\dagger}a_{m\beta} + a_{m\beta}^{\dagger}a_{m\alpha})]^2 = 0 \qquad (9)$$

which yields adiabatic energies by way of the following roots

$$E_{\pm}(q_{i\alpha\beta}) = \tfrac{1}{2}\{\Sigma_{i\alpha\beta}K_{i\alpha\beta}q_{i\alpha\beta}^2 + (E_1 + E_2) \pm \sqrt{[(E_{12})^2 + 4[\Sigma_{i\alpha\beta}G_{i\alpha\beta}q_{i\alpha\beta}]^2]}\} \qquad (10)$$

where $E_{12} = |E_2 - E_1|$ is the fermion states' energy gap. The energies predicted by equation (10) are namely those which play the role of vibronic potentials to form the graphics of Figure 2.

Hamiltonian (8) has been investigated by means of a variational Ansatz [4-6]. Despite the probe wave function being 1D in polaron quantities, it predicts a total-momentum dependence of the polaron coordinate. For a semi-bound polaron, it is a double-maximum curve peaking in the middles of two neighboring quadrants [6].

We had earlier shown that Hamiltonian (7) can be transformed to a form describing rotation in the equatorial (x,y) plane. A constant polar radius $r_\varphi$ is assumed which turns the problem 1D depending on the azimuth angle $\varphi$. This problem is solved in Mathieu's transcendent periodic eigenfunctions of the nonlinear oscillator [7].

## 4. Schrödinger equation for itinerant vibronic polarons

In 3D the Hamiltonian of the eigenvalue equation in cylindrical coordinates (but at constant radius r) is (cf. [8] where the eigenvalue problem is dealt with in spherical coordinates):

$$H_{vib} = -(\hbar^2/2I)\Delta(\varphi,z) \pm (M\omega^2/b)(d_c - d_b)[r_\varphi^4(\cos\varphi)^4 + r_\varphi^4(\sin\varphi)^4 + z^4] + D_\pm \qquad (11)$$

with $\Delta_{\varphi,z} = (1/r)(\partial/\partial r)[r(\partial/\partial r)] + (1/r^2)(\partial^2/\partial\varphi^2) + \partial^2/\partial z^2$.

The problem turns quasi-2D in the equatorial plane (z = 0) (at constant polar radius $r_\varphi$) leading to the eigenvalue equation (where $B_\pm$ and $C_\pm$ are constants) [7]:

$$-(\hbar^2/2I_A)(\partial^2\Phi(\varphi)/\partial\varphi^2) + 2B_\pm\cos(4\varphi)\Phi(\varphi) = (E - C_\pm)\Phi(\varphi) \qquad (12)$$

which is Mathieu's equation. Mathieu's eigenfunctions $\Phi(\varphi)$ describe a quantum-mechanical motion along a circle of radius $r_\varphi$ in (x,y) plane provided $\partial/\partial z \equiv 0$ and $\partial/\partial r \equiv 0$, $\partial^2/\partial r^2 \equiv 0$ which may all hold good in the valley along a reorientational circle. The helical 3D solution should combine rotation in (x,y) plane with an uniform motion by plane-wave along z-axis:

$$\Psi(\varphi,z) = A\exp(ikz)\Phi(\varphi) \qquad (13)$$

The relevant physics applies to the rotation-like reorientation of off-center impurities about their central cation site. If the center is not immobilized but moves uniformly along the z-axis in a plane-wave manner, then the overall 3D motion will be that along a helix. Some time ago we launched this idea for the motion of vibronic polarons composed of translation along z and rotation in equatorial plane about a central site [9]. Going off-center lowers the polaron energy. This is the heart of the matter for it breaks the inversion symmetry to trace a spiral.

The (11) based Schrödinger equation can be split in two independent parts relevant to the above suggestions: one in-plane depending on the azimuth φ in the (equatorial) plane and another one on the coordinate z. Indeed, on setting θ ~ ½π in the sinθ terms we get the azimuth-dependent part of eqn. (12) as well as the z-dependent part of the form:

$$H_{vib,z,\varphi} = -(\hbar^2/2I)\Delta(\theta,\varphi) \pm (M\omega^2/b)(d_c-d_b)[r^4(\cos\varphi)^4 + r^4(\sin\varphi)^4 + z^4] = H_{vib,\varphi} + H_{vib,z}$$

The corresponding complete Schrödinger equation reads

$$H_{vib,z,\varphi} \Psi(z,\varphi) = (E_z + E_\varphi)\Psi(z,\varphi) \qquad (14)$$

with eigenvalue $E = E_\varphi + E_z$ and eigenstate $\Psi(z,\varphi) = \Phi(\varphi)A\exp(ikz)$ according to

$$H_{vib,\varphi} \Phi(\varphi) = E_\varphi \Phi(\varphi)$$

$$H_{vib,z} \exp(ikz) = E_z \exp(ikz) \qquad (15)$$

The angular part has been dealt with in equatorial plane [7]. Its eigenfunctions are Mathieu's periodic functions, while its eigenvalues fall into Mathieu's rotational energy bands [9]. The z-dependent part simplifies to

$$H_{vib,z} = -(\hbar^2/2I)(\partial^2/\partial z^2) \pm (M\omega^2/b)(d_c-d_b)z^4 \; (z \sim 0) \sim -(\hbar^2/2I)(\partial^2/\partial z^2) \qquad (16)$$

which is solved as $A\exp(ikz)$ at small z. Its corresponding eigenvalue is given by $E_z = \hbar^2 k^2/2I$. At medium and large z the complete Hamiltonian (14) describes two kinds of anharmonic vibrators, one with a positive curvature, the other one with negative curvature. They are both solvable. The vibronic potentials arising from the anharmonic parabolas are seen in Figure 3.

### 5. Global symmetry-breaking mechanism

Herein we will paraphrase the statement appropriate for the Jahn-Teller effects [10] to make it adaptable to other choices of fermions and boson fields: For any pair of (nearly)-degenerate (opposite)-parity fermion states there will be a boson field of appropriate spatial symmetry which couples to the fermions to mix their states thus generating a hybrid state which lifts the (near)-degeneracy by lowering the symmetry. (Bracketed words are alternatives.) This statement will have to be refined in a further publication.

### 6. Implications for light propagation in material medium

The above statements may come to solving the controversy of light propagation in that the vibronic coupling transforms the rectilinear translational motion into one along a spiral. This is a profound conclusion for ascertaining that the Jahn-Teller effects are likely to be the universal driving force behind the breaking of symmetry in nature!

A vibronic equation describes the behavior of the boson wave function reacting to the mixing of fermion states by the boson-to-fermion coupling. If quanta of light (photons) are involved in their capacity of a quantized boson field, then the fermion-boson coupling may arise in a medium of electron or hole states.

Alternatively, fermion states and coupled boson fields may be found in nuclear matter as well in the form of nucleonic fermions and coupled γ-quanta or coupled π-mesons. In all these cases the resulting average vibronic trajectory of the coupled bosons will be helical. If beyond the present-day capacity for an experimental verification due to tiny curvatures [1], yet the theoretical model based on the breaking of inversion symmetry through vibronic coupling is worth advancing for it leads to far reaching consequences. In particular, it raises the questions as to the form and physical space of light propagation which seem to be interrelated for if the form is helical, the space is 3D [1].

An important quantity whose numerical value is of utmost importance for the experimental verification of the rectilinear-to-helical path interconversion is the magnitude of the spiral (polar) radius $r_\varphi$. Kolyvodiakos has estimated it from equating the gravitational attraction between two photons with the centrifugal force on rotating along the spiral which resulted in a desperately tiny numerical value, the order of $10^{-76}$ cm, the reason being the extremely small photon mass assumed [1]. For off-center defects such as substitutional $Li^+$ ions in KCl, the spiral radius is given by the off-center displacement which is the order of a few tenths of an Ångström unit [11]. This being the order of magnitude for the spiral radius of a vibronic polaron too, the latter would have all the chances to be observed experimentally. It will be very interesting to see just how large the spiral radii are in nuclear matter too.

The estimates of the spiral radii tells directly of the nature of the symmetry breaking involved in the rectilinear-to-helical transformation: Namely, the off-center displacement occurring as a result of the breaking up of inversion symmetry, its also serving as a helical radius would imply that the spiral too breaks the inversion symmetry. This conclusion is important, for it helps identify the intimate details of the relevant symmetry breaking. On the other hand, the similarities between the assumed light and vibronic polaron propagation makes it possible to use the latter as a model for collecting data that can eventually benefit studying the former.

Acknowledgement. We gratefully acknowledge our using excerpts of an Internet-based survey on helices [2] for an appropriate background to our subsequent discussion.

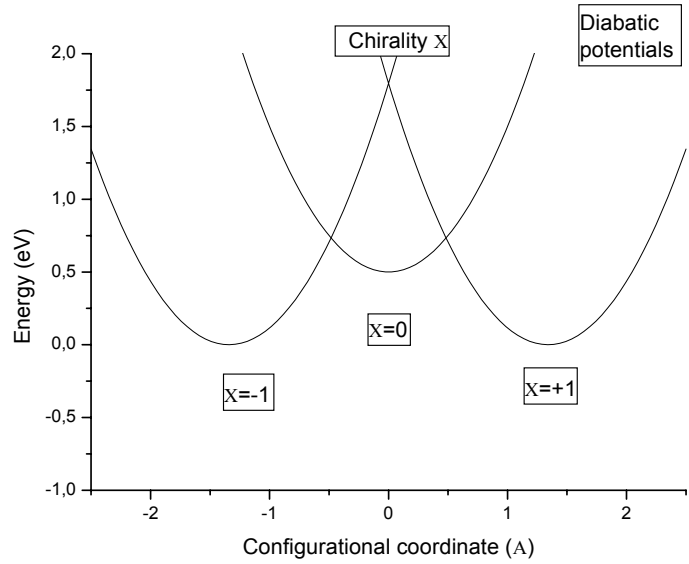

Figure 1: Diabatic potentials (no crossover splitting) illustrating three conceivable cases of handedness: left-handed (X=-1), right-handed (X=+1), and no-handed (X=0).

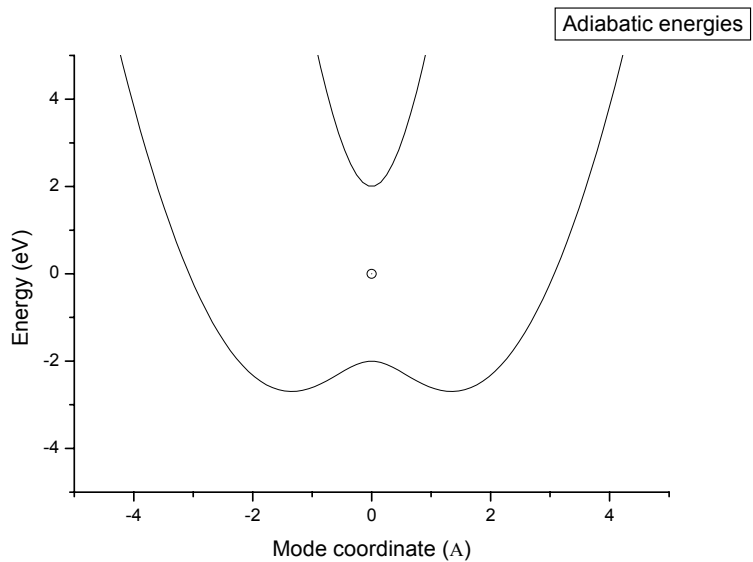

Figure 2: Adiabatic energies (finite crossover splitting) illustrating the obtained vibronic potentials following equation (10). The parameters used for the calculation are: G = 3 eV/Å, K = 2 eV/Å$^2$, $E_{12}$ = 4 eV, $E_{JT}$ = 2.25 eV.

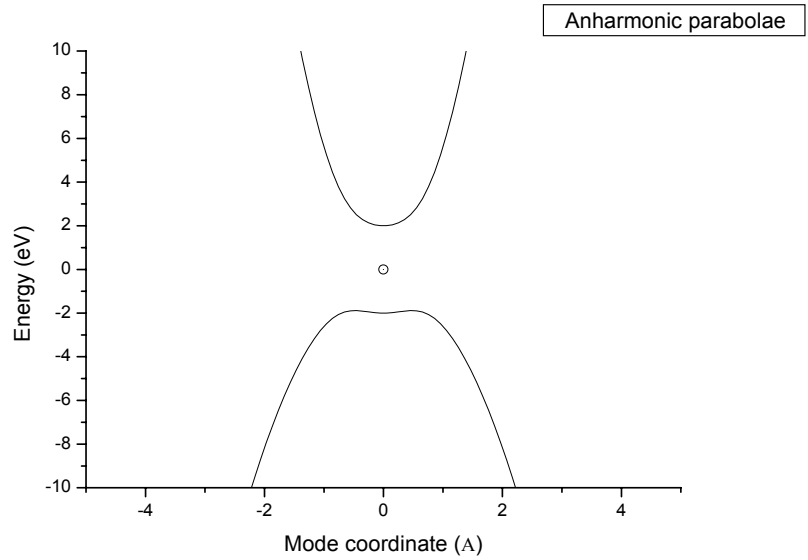

Figure 3: Adiabatic potentials based on a $z^4$- power parabola entering the z-component vibronic equation (16). At small z the potential is close to constant which justifies the presumed approximations. The parameters used for the calculation are those of the preceding Figure 2. The lower parabola would not favor the formation of a bound state to trap the itinerant polaron, as assumed.